\documentclass[9pt,conference]{IEEEtran}
\IEEEoverridecommandlockouts

\usepackage{cite}
\usepackage{amsmath,amssymb,amsfonts}
\usepackage{graphicx}
\usepackage{bm}
\usepackage{booktabs} 
\usepackage{diagbox} 
\usepackage{enumitem}
\usepackage{multicol}
\usepackage{multirow}
\usepackage{makecell}
\usepackage{pifont}
\usepackage{subfigure}
\usepackage{xcolor}
\usepackage[linesnumbered,lined,ruled,commentsnumbered]{algorithm2e}
\usepackage{url,hyperref,cleveref}

\def\BibTeX{{\rm B\kern-.05em{\sc i\kern-.025em b}\kern-.08em
    T\kern-.1667em\lower.7ex\hbox{E}\kern-.125emX}}

\DeclareMathOperator*{\argmax}{arg\,max}

\begin{document}

\title{NTC-KWS: Noise-aware CTC for \\ Robust Keyword Spotting}

\author{\IEEEauthorblockN{Yu Xi$^1$, Haoyu Li$^1$, Hao Li$^2$, Jiaqi Guo$^2$,  Xu Li$^2$, Wen Ding$^3$, Kai Yu$^{1{\dagger}}$ \thanks {$^{\dagger}$ Corresponding Author.}}
\IEEEauthorblockA{\textit{$^1$MoE Key Lab of Artificial Intelligence, AI Institute, X-LANCE Lab, Shanghai Jiao Tong University, Shanghai, China}}
\IEEEauthorblockA{\textit{$^2$AISpeech Ltd, Suzhou, China}~~~~~\textit{$^3$NVIDIA, Shanghai, China}}
\IEEEauthorblockA{
\{yuxi.cs, haoyu.li.cs, kai.yu\}@sjtu.edu.cn~~~ \{hao.li, jiaqi.guo, xu.li\}@aispeech.com~~~ wend@nvidia.com
}
}
\maketitle

\begin{abstract}
In recent years, there has been a growing interest in designing small-footprint yet effective Connectionist Temporal Classification based keyword spotting (CTC-KWS) systems. They are typically deployed on low-resource computing platforms, where limitations on model size and computational capacity create bottlenecks under complicated acoustic scenarios. Such constraints often result in overfitting and confusion between keywords and background noise, leading to high false alarms. To address these issues, we propose a noise-aware CTC-based KWS (NTC-KWS) framework designed to enhance model robustness in noisy environments, particularly under extremely low signal-to-noise ratios. Our approach introduces two additional noise-modeling wildcard arcs into the training and decoding processes based on weighted finite state transducer (WFST) graphs: self-loop arcs to address noise insertion errors and bypass arcs to handle masking and interference caused by excessive noise. Experiments on clean and noisy Hey Snips show that NTC-KWS outperforms state-of-the-art (SOTA) end-to-end systems and CTC-KWS baselines across various acoustic conditions, with particularly strong performance in low SNR scenarios.
\end{abstract}


\section{Introduction}
Keyword spotting (KWS), also referred to as wake word detection (WWD), serves as a vital interface for human-machine interaction~\cite{icassp2014-guoguochen-dnn_kws,streaming-kws-2,streaming-kws-1,deep-spoken-kws-overview}. KWS systems are typically deployed on devices and operate continuously. As application scenarios grow more diverse, improving noise robustness under resource-constrained conditions has become a key research focus in the field of KWS.

Various approaches have been developed to enhance the noise robustness of KWS systems. One approach is to introduce single/multi-channel speech enhancement (SE) modules preceding the KWS module~\cite{is2018-mengyu-text_dependent_se_for_kws,is2020-mengyu-e2e_multi-look_kws,iconip2020-xueliangzhang-joint-se-kws,icassp2022-yueyuena-multilook_se_and_multilook_kws_joint_egonoise_suppression_and_kws_on_sweeping_robots,is2023-chouchangyang-SE_speech_presence_probability,is2024-haoyuli-tpdt_kws}. Another approach involves designing new training strategies or architectures to achieve noise-robust KWS systems~\cite{taslp2021-Lopez_Espejo-loss_and_strategy_for_noise_robust_kws,icssp2024-tdt-kws,icassp2024-haizhu-TCSN_for_noisy_kws}. 
Regardless of the methods, simulated noisy data plays a critical role in enhancing KWS performance. Our preliminary experiments highlight two key observations: First, at low signal-to-noise ratio (SNR) levels, strong noise energy naturally leads to lower recall; Second, simulated data with excessively low SNR causes the model to overfit to noise, utimately degrading overall performance. These occur because the noise in challenging training data masks portions of keyword speech, leading the model to confuse noise with the target keyword. 


Noise-induced speech errors can lead to mismatches between the speech content and paired text transcripts during training. Previous CTC-based ASR approaches have introduced wildcard modeling training method~\cite{iclr2022-xingyucai-wctc_firstwork_of_ctc_for_incompelte_labels,neurips2022-vinellPratap-STC_arbitrary_number_missing_with_indeterminate_location,is2023-dongjigao-bypass_temporal_classification,asru2023-dongjigao-omni_temporal_classification} to handle transcript inaccuracies. Conversely, this study addresses noise-induced speech errors and proposes the \textit{NTC-KWS} framework, which integrates both noise-aware training and decoding strategies to enhance model robustness in noisy conditions. More specifically, we incorporate two types of noise-modeling wildcard arcs into the CTC training framework to adaptively model noise and further improve the WFST-based decoding search space, boosting KWS performance.

The contributions of this work are summarized as follows:
\begin{enumerate}
    \item We propose noise-aware CTC training criterion to model noise-dominant speech segments under complex acoustic scenarios to prevent noise overfitting. 
    \item We are the first to adopt a training-free decoding strategy that incorporates noise-modeling paths into WFST-based KWS decoding. This approach enhances performance under noisy environments without requiring NTC training and offers further improvements when applied within our NTC-KWS framework.
    \item Compared to SOTA end-to-end and CTC-KWS baselines, the proposed NTC-KWS achieves superior performance across both clean and noisy conditions, with notable improvements in extremely challenging scenarios.
\end{enumerate}

\section{Methodology}

\begin{figure*}[t]
\vspace{0em}
\centerline{\includegraphics[width=17.5cm]{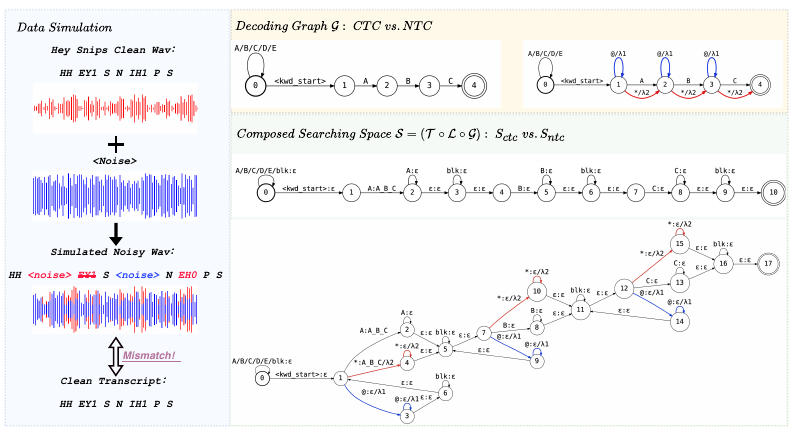}}
\caption{The overview illustrates noisy data simulation and comparisons of CTC and NTC WFST-based decoding graphs. In the left part, we show three types of errors by a noise simulation example introduced by overwhelming noise: masking and interference (shown in red), and insertion (shown in blue). In the right part, we provide a toy example where the keyword is set to ``A B C". The decoding transition rules under grammar and token levels are represented by $\mathcal{G}$ and $\mathcal{S}$, respectively. In NTC, we highlight two types of additional wildcard arcs in red and blue, corresponding to the error colors of the simulated example shown on the left. In the composed graphs $\mathcal{S}$, $\lambda_{1}$ and $\lambda_{2}$ represent the wildcard transition costs. $\epsilon$ to the left of $:$ denotes an empty transition, while on the right, it indicates a null output.}
\label{fig:overview_figure}
\end{figure*}	

\subsection{WFST-based Implementation of CTC}
\label{sec:wfst-ctc}
Connectionist temporal classification (CTC)~\cite{icml2006-alex_graves-CTC} is a widely used loss function for training sequence learning model with length-mismatch of inputs and labels, and is naturally suitable for ASR or KWS. For ASR, an input acoustic feature sequence can be represented as $\mathbf{x} = [x_{1},x_{2},...,x_{T}] \in \mathbb{R}^{T \times D}$, where each $x_{t}$ is a $D$-dimensional acoustic feature vector and $T$ represents the total frames. The output label sequence can be represented as $\mathbf{y} = [y_{1},y_{2},...,y_{U}] \in \mathbb{R}^{U \times 1}$, where $y_{u}$ is the label token and $U$ is the number of total tokens. It's worth noting that $T$ is must larger than $U$ because CTC introduces a special blank token $\phi$ to model the null output at each frame. The actual alignment sequence for the input is $\bm{\pi} = [\pi_{1},\pi_{2},...,\pi_{T}] \in \mathbb{R}^{T \times 1}$, where each $\pi_{t}$ represents either a normal phoneme or a blank token $\phi$ in the CTC vocabulary $\mathcal{V}_{ctc}$. 
The objective of CTC is to minimize the summation of the negative log-likelihood of the alignment sequences given the input sequence. Therefore, the CTC loss can be formulated as:
\begin{equation}
    \mathcal{L}_{CTC} = - \log P(\mathbf{y}|\mathbf{x}) = - \log \sum_{\bm{\pi} \in \mathcal{B}^{-1}(\mathbf{y})} P(\bm{\pi}|\mathbf{x}),
\end{equation}
where $\mathcal{B}$ is defined as a mapping from the legal CTC alignment $\bm{\pi}$ to the label $\mathbf{y}$ and $\mathcal{B}^{-1}$ stands for the inverse mapping. In addition, the decoding is to find the most possible target of the speech. We can formulate it as:
\begin{equation}
\label{eq:eq_decoding_2} 
    \mathbf{\hat{y}} = \argmax_{\mathbf{y}} P(\mathbf{y}|\mathbf{x}) \approx \argmax_{\mathbf{y}, \bm{\pi} \in \mathcal{B}^{-1}(\mathbf{y})} P(\bm{\pi}|\mathbf{x}) ,
\end{equation}
where the approximate derivation introduces the Viterbi decoding to find the most possible alignments.

Weighted finite state transducer (WFST)~\cite{Computer_Speech_Language2002-mohri-wfst_for_asr}, aiming to map an input sequence to an output one, provides an efficient implementation for the CTC training and decoding. Specifically, the searching space $\mathcal{S}$ is defined by the composition operations of $\mathcal{T}$, $\mathcal{L}$ and $\mathcal{G}$, which represent CTC merging (remove $\phi$ and merge repeated tokens), lexicon (a mapping from word to tokens) and grammar FSTs, respectively. In addition, the emission dense graph $\mathcal{E}$ is built to model acoustic posteriors. So we can reformulate the CTC training and decoding as follows: 
\begin{equation}
    \mathcal{L}_{CTC} = - \textit{Forward}(\;\mathcal{E}(\mathbf{x}) \cap (\;\mathcal{T} \circ \mathcal{L} \circ \mathcal{G}(\mathbf{y})\;)\;),
\end{equation}
and
\begin{equation}
    \mathbf{\hat{y}} = \textit{Viterbi}(\;\mathcal{E}(\mathbf{x}) \cap (\;\mathcal{T} \circ \mathcal{L} \circ \mathcal{G}\;)\;).
\end{equation}
Here, $\cap$ and $\circ$ denote the intersection and composition operations on WFSTs, respectively. $\mathcal{G}(\mathbf{y})$ represents a linear FST constructed for the transcript $\mathbf{y}$. The terms~\textit{Forward} and~\textit{Viterbi} refer to the forward computation and Viterbi decoding on the composed WFST graph. Specifically,~\textit{Token Passing} is an efficient method for implementing Viterbi decoding within the WFST framework. When a token reaches the final node, a decoding path is considered complete.

\subsection{CTC-based Keyword Spotting}
Compared to the CTC-based ASR described in~\Cref{sec:wfst-ctc}, KWS follows a nearly identical training procedure, with only two minor differences in the decoding phase.

\textbf{Decoding graph.}
The grammar graph $\mathcal{G}$ of the KWS~\cite{hmm_filler_2_no_others,asru2017-yanzhanghe-classic_ctc_rnnt_kws,arxiv2020-theodore-quantized_lstm_ctc_kws,icassp2024-yuxi-contrasitve_learning_for_kws} is more compact and lightweight. It consists of only keyword paths and an optional background path rather than the whole acoustic space. 

\textbf{Confidence score measure.} Search results are transformed into keyword activation scores by computing confidence scores of speech segments. Efforts have been made to evaluate different strategies for computing confidence scores for CTC-based KWS~\cite{asru2017-yanzhanghe-classic_ctc_rnnt_kws,arxiv2020-theodore-quantized_lstm_ctc_kws}. In this work, we follow the confidence metric proposed in~\cite{arxiv2020-theodore-quantized_lstm_ctc_kws}.

\subsection{Training and Decoding of NTC-KWS}

Although CTC-based KWS performs well in typical acoustic conditions, its performance degrades considerably when speech segments are overwhelmed by noise. As shown on the left side of~\Cref{fig:overview_figure}, excessive noise introduces errors in the simulated speech, including content masking, noise artifacts, and pronunciation distortion. This results in a mismatch between the speech and the transcript, increasing the possibility of noise-induced false triggers.
To enhance noise robustness, we propose NTC-KWS, which introduces two types of wildcard arcs: self-loop arcs and bypass arcs during both training and decoding. These arcs represent distinct token-level graph errors in $\mathcal{G}$, as shown in~\Cref{fig:overview_figure}. When parts of the keyword are obscured by noise, self-loop arcs (\texttt{@}) model insertion errors in keyword labels (denoted by blue errors and arcs). In scenarios where phonemes are distorted or masked by noise, bypass arcs (\texttt{*}) account for substitution or deletion errors (depicted as red errors and arcs). These wildcards effectively capture the concept of ``\textbf{excessive speech noise and interference}". The acoustic posteriors for \texttt{*} and \texttt{@} arcs, representing \textbf{the average posteriors across all speech}, are computed as follows:
\begin{equation}
    \begin{aligned}
       P(\pi_{i}(\texttt{@})|x_{i}) =  P(\pi_{i}(\texttt{*})|x_{i}) =  \frac{\sum_{j\in\mathcal{V}_{ctc}/\{\phi\}}P(\pi_{i}(j)|x_{i})}{len(\mathcal{V}_{ctc}/\{\phi\})},
   \end{aligned}
\end{equation}
where $\pi_{i}(j)$ denotes the predicted token $j$ in vocabulary $\mathcal{V}_{ctc}$ at frame $x_{i}$. To ensure that the model prioritizes learning the correct alignments over the wildcard paths, large initial penalties are assigned to the wildcard arcs. These penalties decay exponentially with respect to the training epochs:
\begin{equation}
\lambda^{(l)}_{\texttt{@}} = \lambda^{(0)}_{\texttt{@}} * \beta_{\texttt{@}}^{l},\quad \lambda^{(l)}_{\texttt{*}} = \lambda^{(0)}_{\texttt{*}} * \beta_{\texttt{*}}^{l},
\end{equation}
where $l$-th denotes the number of the current epoch, and $\lambda^{(0)}_{\texttt{@}}$, $\beta_{\texttt{@}}$, $\lambda^{(0)}_{\texttt{*}}$, $\beta_{\texttt{*}}$ are hyper-parameters, set empirically.

Although the model has learned to handle noise-dominant parts of the keyword using wildcard paths through NTC-based training, the standard CTC-KWS graph for decoding fails to fully leverage the enhanced training, especially in low SNR scenarios. To better alignment the training and decoding procedure, we must also expand the decoding search space with wildcard paths by modifying the grammar graph $\mathcal{G}$, as shown in the right part of~\Cref{fig:overview_figure}. For example, assuming the keyword is ``A B C" and the vocabulary $\mathcal{V}_{ctc}$ includes the tokens ``A B C D E" along with the blank token $\phi$, we introduce self-loop (\texttt{@}) and bypass (\texttt{*}) arcs between adjacent sub-words of the ``A B C" keyword sequence on graph $\mathcal{G}$ to model potential errors accordingly. The NTC grammar graph $\mathcal{G}_{ntc}$ is then composed with the transition $\mathcal{T}$ and lexicon $\mathcal{L}$ to create the NTC-KWS search space $\mathcal{S}_{ntc}$. Compared to the CTC-KWS search space $\mathcal{S}_{ctc}$, $\mathcal{S}_{ntc}$ offers alternative noise-modeling paths for token passing, which improves decoding topology for noisy KWS in the following aspects. First, the alignment of training and decoding methods ensures consistency. Second, recognizing every token in the keyword becomes challenging when noise is predominant. Therefore, loosening decoding restrictions in complex acoustic environments is reasonable to boost recall, as it gives the model the option to select wildcard arcs against noise. To balance recall and false alarms, we introduce transition costs $\lambda_{\texttt{@}}$ and $\lambda_{\texttt{*}}$ ($\lambda_{1}$ and $\lambda_{2}$ in~\Cref{fig:overview_figure}) for the two types of arcs in the NTC decoding graph, respectively. However, if a decoding path consists solely of wildcard tokens without containing any keyword phonemes, that path will be discarded.


\section{Experimental Setup}

\subsection{Datasets}
\label{context:dataset}
To assess the performance and noise robustness of the systems, we construct datasets using the following open-source corpora.

\begin{itemize}
    \item \textbf{LibriSpeech (LS)}~\cite{LibriSpeech}: LibriSpeech is an open-source, widely used ASR corpus containing 960 hours of English speech with corresponding transcripts. The dataset is used as the pre-training dataset for the base model and general ASR data for KWS models.
    \item \textbf{Hey Snips (Snips)}~\cite{snips}: Hey-Snips is an open-source dataset for KWS centered around the keyword ``Hey Snips." Due to the absence of transcripts for negative utterances, the negative data is used exclusively for evaluation and not for training. We combined all negative data from the train, dev, and test sets to create a large test set, following the data preparation approach in~\cite{icssp2024-tdt-kws}. The duration of the new negative test set is about 97 hours.
    \item \textbf{WHAM!}~\cite{wham}: The WHAM! dataset consists of various types of ambient noise, including music, restaurant sounds, etc. We incorporate this dataset to synthesize noisy utterances for both training and evaluation. 
\end{itemize}

\textbf{Training.} We pre-train the base speech encoder using clean LibriSpeech. Then both Snips and an equivalent amount of LibriSpeech are utilized to fine-tune a KWS model. To simulate noisy dataset, each clean waveform from Snips and LS is mixed with a randomly selected noise sample from WHAM!, with the SNR chosen from a uniform distribution ranging from 0 to 20 dB. The clean and noisy speech are gathered to construct the training dataset.

\textbf{Evaluation.} 
We simulate noisy negative test set in the same manner as the training setting and simulate test positive speech at different fixed SNR levels. All positive test samples from the Snips dataset are mixed with noise at SNR levels of $\{0, 5, 10, 15, 20\}$ dB. To further evaluate robustness, we simulate an additional dataset under -5 dB, representing an extremely low and out-of-domain SNR.

\subsection{Configuration}

\textbf{Models.} We leverage the toolkit K2~\cite{k2} to build the differential WFSTs for training and leverge OpenFST~\cite{naacl2009-michael_riley-openfst} to build decoding systems. The Deep Feedforward Sequential Memory Network (DFSMN)~\cite{DFSMN}, a lightweight yet effective speech encoder, is a mainstream architecture for KWS tasks~\cite{icassp2021-yuekaizhang-tinyrnnt_kws_Tencent,icassp2021-yaotian-rnnt_fsmn_160khrs_kws_ByteDance,icssp2024-tdt-kws}. We use this architecture as the backbones for both CTC and NTC KWS systems. All models are initialized from the same pre-trained checkpoint and optimized by SGD~\cite{SGD}. The DFSMN architecture consists of 6 layers, with hidden and projection sizes set to 512 and 320, respectively. Each DFSMN layer has left and right context orders of 8 and 2. The final output unit of the CTC-based model includes 70 monophones derived from the CMU Pronouncing Dictionary ``cmudict-0.7b"~\cite{cmudict}, along with the special blank symbol $\phi$. Additionally, the NTC includes up to two extra special tokens: the self-loop token \texttt{@} and the bypass token \texttt{*}.

\textbf{Acoustic features.} We extract 40-dimensional log Mel-filter bank coefficients (FBank) from the raw audio using a 25ms window and a 10ms hop. Additionally, we employ two data augmentation strategies: online speech perturbation~\cite{Speed_Perturbation}, with a warping factor uniformly sampled from $\{0.9, 1.0, 1.1\}$, and SpecAugment~\cite{Specaug}, using 2 frequency masks with a maximum $F=10$ and 2 time masks with a maximum $T=50$ for each utterance. We concatenate the current frame with 5 preceding and 5 succeeding frames to create the model input. We also apply a frame-skipping factor of 3 to reduce computational cost through down-sampling.

\textbf{Hyper-parameters.}
Decay coefficients $\beta_{\texttt{@}}$ and $\beta_{\texttt{*}}$ are set to 0.999 and 0.975, respectively. The initial wildcard costs $\lambda^{(0)}_{\texttt{@}}$ and $\lambda^{(0)}_{\texttt{*}}$ are both set to -4 for training. Decoding coefficients $\lambda_{\texttt{@}}$ and $\lambda_{\texttt{*}}$ are tuned during test. There are at most 20 active tokens during token passing inference for both CTC and NTC systems.


\subsection{Evaluation Details}
\textbf{Baselines.} We construct comprehensive baselines to evaluate our NTC-KWS system. Firstly, we compare the CTC-based model with the state-of-the-art (SOTA) end-to-end (E2E) systems for the Hey Snips dataset, which is open-sourced in WeKws~\cite{icassp2023-binbinzhang-wekws}, to prove the KWS models constructed by CTC or its variants can get superior performance. Then we further conduct detailed comparison among NTC-KWS and CTC-related baselines at different in-domain and out-of-domain SNRs to evaluate the improved training and decoding methods.

\textbf{Evaluation metrics.} We report recall values at various SNR levels for a specific false alarm rate (FAR). Published SOTA models typically present recalls at FAR levels of 0.5 or 1 per hour. To ensure a fair comparison, we adopt the same settings when benchmarking against these systems. However, this evaluation standard is relatively lenient, as performance tends to be strong at such FAR levels. To minimize the impact of deviations and randomness, we assess all systems under stricter conditions. Therefore, after the initial comparison, we present recall at FAR = 0.05 per hour.

\section{Results and Analysis}

\subsection{The Comparison with E2E SOTA Baselines}

\begin{table}[t]
  \centering
  \newcolumntype{S}{>{\small}c}
  \linespread{0.9}
  \caption{The comparisons of E2E baselines and our CTC and NTC KWS systems on the clean Hey Snips dataset. ``Pos. Snips" refers to the positive portion of the Snips dataset, while ``Equ. LS" indicates that we add an equal amount of LibriSpeech utterances as general ASR data.}
  \begin{resizebox}{1.0\columnwidth}{!}
  {
    \begin{tabular}{c| c| c c| c c c}
      \toprule
    \multirow{2}*{\textbf{ID}} & \multirow{2}*{\textbf{System}} &  \multirow{2}*{\textbf{\makecell{Training \\ Data}}} & \multirow{2}*{\textbf{\makecell{Test Neg. \\ Duration}}} & \multicolumn{3}{c}{\textbf{FAR}} \\
      \cmidrule(lr){5-7}
     ~ & ~ & ~ & ~ & 0.05 & 0.5 & 1.0 \\
      \midrule
     A& RIL KWS~\cite{is2020-kunzhang-first_baseline_of_snips_in_wekws} & \multirow{3}*{\makecell{Official\\ Snips}} &  \multirow{3}*{23hrs} & - & 96.47 & 97.18 \\ 
     B& WaveNet~\cite{icassp2019-Coucke-second_baseline_of_snips_in_wekws} &  ~ &  ~ & - & 99.88 & - \\
     C1& WeKws-MDTC~\cite{icassp2023-binbinzhang-wekws} &  ~ & ~ & - & 99.88 & 99.92 \\
      \midrule
     C2& WeKws-MDTC & \multirow{3}*{\makecell{Pos. Snips \\ + Equ. LS}} &  \multirow{3}*{97hrs} & 89.52 & 98.85 & 99.29 \\
     D & CTC-KWS & ~ & ~ & 98.77 & \textbf{99.72} & \textbf{99.76} \\
     E& NTC-KWS & ~ & ~ & \textbf{98.97} & 99.64 & 99.72 \\
      \bottomrule
    \end{tabular}%
    }
   \end{resizebox}
  \label{table:sec4.1}
\end{table}
The upper part of the table shows the results using the official Snips data, while the bottom part reports results based on reconstructed data due to the lack of transcriptions for the Snips negative samples. This reconstruction is necessary since CTC-based models cannot be trained on speech data without transcripts, as discussed in~\Cref{context:dataset}. We compare our CTC and NTC models with SOTA E2E baselines on the clean Hey Snips dataset at FAR of 0.5 or 1.0 per hour. Our models can achieve comparable performance (D\&E vs. B or C1). When models are trained on the same data, the results in the bottom section of~\Cref{table:sec4.1} show that both CTC and NTC models outperform the SOTA baseline at FARs of 0.5 or 1.0 per hour (D\&E vs. C2). Additionally, under the more challenging condition of FAR = 0.05 per hour, CTC and NTC models exhibit an absolute recall improvement of over 9\%. The results in~\Cref{table:sec4.1} further demonstrate that our models surpass the baselines, especially under strict test conditions, validating the effectiveness of the CTC-KWS baseline and proposed NTC-KWS.

\subsection{The Impact of NTC Training and Decoding}

\begin{table}[b]
  \centering
  \vspace{-10pt}
  \newcolumntype{S}{>{\small}c}
  \linespread{0.9}
  \caption{The evaluation of the different combinations for CTC-KWS and NTC-KWS. ``Train" refers to the specific training criterion employed during the model training phase, while ``Test" indicates the type of decoding graph used during the inference phase. }
    \begin{resizebox}{1.0\columnwidth}{!}
    {
    \begin{tabular}{c| c c | c | c c c c c |c}
      \toprule
      \multirow{2}*{\textbf{ID}} &\multirow{2}*{\textbf{Train}} & \multirow{2}*{\textbf{Test}} & \multicolumn{6}{c|}{\textbf{SNR}} & \multirow{2}*{\textbf{Avg.}} \\
      \cmidrule(lr){4-9}
      ~ &~ & ~ & -5 & 0 & 5 & 10 & 15 & 20 & ~ \\
      \midrule
        D & CTC & CTC & 41.6 & 75.1 & 90.2 & 95.8 & 97.5 & 98.3 & 83.1 \\ 
        F & NTC & CTC & 41.7 & 74.3 & 89.8 & 95.8 & 97.5 & 98.1 & 82.9 \\
        G & CTC & NTC & 52.0 & 79.8 & 91.7 & 96.2 & 97.4 & 98.0 & 85.8 \\
        E & NTC & NTC & \textbf{56.6} & \textbf{83.4} & \textbf{93.2} & \textbf{97.6} & \textbf{98.2} & \textbf{98.7} & \textbf{88.0} \\
      \bottomrule
    \end{tabular}
   }
   \end{resizebox}
  \label{table:sec4.2}
\end{table}

In this section, we examine the individual effects of NTC training and decoding. \Cref{table:sec4.2} illustrates the performance across different combinations of training and inference strategies for CTC and NTC. Given the CTC inference (F vs. D), updating the training criterion to NTC alone does not improve performance, as NTC training does not compel the model to treat noise as part of the speech signal. As a result, model F behaves similarly to a CTC model trained on clean, high-quality speech. Training the model with CTC loss and testing it using the NTC decoding method (D vs. G) yields a 2.7\% absolute improvement in average recall. This gain is particularly pronounced at low SNR levels (SNR = -5, 0), highlighting the challenge of token decoding under such conditions. Our proposed decoding strategy relaxes constraints in complex environments, leading to significant performance gains. This suggests that if retraining the original CTC KWS model is not feasible but performance improvements in noisy settings are required, adopting the NTC decoding strategy offers a practical solution. Furthermore, applying NTC decoding to the NTC model results in a 2.2\% absolute increase in average recall (E vs. G), primarily due to the alignment between the training and inference processes. This improvement highlights the benefits of consistency between NTC training and decoding. In summary, the proposed NTC-KWS framework achieves a 4.9\% absolute improvement over the CTC-KWS baseline (E vs. D). Notably, there is a significant performance gain of 15\% and 8.3\% at low SNRs of -5 and 0, respectively, underscoring the robustness of NTC-KWS in extremely noisy conditions.

\subsection{Search for Optimal Decoding Configuration}
This section further investigates the importance of two hyper-parameters in the decoding configuration. 
Grid search is performed on the values of $\lambda_{\texttt{@}}$ and $\lambda_{\texttt{*}}$. The best performance is achieved when both $\lambda_{\texttt{@}}$ and $\lambda_{\texttt{*}}$ are set to positive values, indicating that the NTC model does not overfit to shortcuts during NTC training and still requires boosting paths of wildcards during inference. We also experimented with setting the decoding transition coefficients to negative values. While the improvement is less pronounced than with positive values, it still outperforms the CTC-KWS baseline. When both coefficients are set to -$\infty$, the NTC-KWS model behaves like model F in~\Cref{table:sec4.2}, indicating that the NTC decoding strategy collapses into the CTC approach. Another finding from~\Cref{table:sec4.3} shows that the decoding phase is more sensitive to $\lambda_{\texttt{*}}$ than $\lambda_{\texttt{@}}$, and setting $\lambda_{\texttt{*}}$ too high cannot be feasible. In future work, we plan to conduct experiments across various noise types to further investigate these decoding parameters.

\begin{table}[h]
    \centering
    \newcolumntype{S}{>{\small}c}
    \linespread{0.9}
    \caption{Different configurations of self-loop weight $\lambda_{\texttt{@}}$ and bypass weight $\lambda_{\texttt{*}}$ during inference. Negative numbers mean punishment, while positive numbers mean encouragement.}
    \begin{resizebox}{1.0\columnwidth}{!} {
    \begin{tabular}{ c c | c | c c c c c |c}
        \toprule
        \multirow{2}*{\textbf{$\lambda_{\texttt{@}}$}} & \multirow{2}*{\textbf{$\lambda_{\texttt{*}}$}} & \multicolumn{6}{c|}{\textbf{SNR}} & \multirow{2}*{\textbf{Avg.}} \\
        \cmidrule(lr){3-8}
        ~ & ~ & -5 & 0 & 5 & 10 & 15 & 20 & ~ \\
        \midrule
        4 & 2 & 56.6 & 83.4 & 93.2 & 97.6 & 98.2 & 98.7 & 88.0 \\
        \midrule
        4 & 4 & 40.0 & 70.9 & 88.8 & 94.9 & 96.9 & 97.4 & 81.5 \\
        4 & 0 & 55.2 & 82.3 & 92.8 & 97.2 & 97.9 & 98.4 & 87.3 \\
        4 & -$\infty$ & 55.0 & 82.1 & 92.8 & 96.9 & 97.9 & 98.3 & 87.2 \\
        \midrule
        2 & 2 & 50.3 & 81.7 & 92.4 & 97.0 & 98.1 & 98.7 & 86.4 \\
        0 & 2 & 49.5 & 80.9 & 92.4 & 96.9 & 98.0 & 98.6 & 86.0 \\
        -$\infty$ & 2 & 49.3 & 80.9 & 92.4 & 96.8 & 98.1 & 98.7 & 86.0 \\
        \bottomrule
    \end{tabular}
    } \end{resizebox}
    \label{table:sec4.3}
\end{table}

\section{Conclusion}
In this paper, we present NTC-KWS, a variant of the CTC-based KWS framework tailored for noisy environments, particularly in extremely low SNR conditions. By incorporating various types of wildcards during both the training and decoding phases, we reduce the risk of model overfitting to noise. Modeling noise-dominant segments as wildcard tokens enables the model to better distinguish keyword speech from background noise. Our system achieves SOTA performance on the clean Hey Snips dataset compared to previous E2E systems. Specifically, the NTC-KWS framework demonstrates an absolute improvement of 4.9\% in average recall across all SNR levels over the standard CTC-KWS system. In low SNR conditions, such as -5 dB and 0 dB, it achieves absolute recall gains of 15\% and 8.3\%, respectively. These improvements underscore the robustness of the proposed system in challenging, noisy environments.

\bibliographystyle{style/IEEEtran}
\bibliography{citations/refs}

\end{document}